\documentclass{PoS}
\newcommand{\be}{\begin{equation}}
\newcommand{\ee}{\end{equation}}
\title{Measuring the trilinear neutral Higgs boson couplings
 in the MSSM at $e^+ e^-$  colliders}

\ShortTitle{MSSM trilinear Higgs couplings}

\author{\speaker{Charanjit K. Khosa}\\%\thanks{A footnote may follow.}
        Centre for High Energy Physics, Indian Institute of
Science, Bangalore 560 012, India\\
        E-mail: \email{khosacharanjit@gmail.com}}

\author{P. N. Pandita\\
        Centre for High Energy Physics, Indian Institute of
Science, Bangalore 560 012, India\\
       E-mail: \email{pandita@associates.iucaa.in}}

\abstract{ We consider the measurement of the trilinear couplings 
 of the neutral Higgs bosons~($H^0, h^0$) in the minimal 
supersymmetric standard model~(MSSM) at a high energy 
$e^+$ $e^-$ linear collider in the light of the discovery 
of a Higgs boson at the CERN  Large Hadron Collider~(LHC). We 
identify the state observed at the LHC with the lightest 
CP-even Higgs boson of the MSSM. We implement this constraint, 
as well as all the other relevant experimental constraints, 
on the parameter space of the MSSM in order to study the 
feasibility of measuring the trilinear couplings of the 
neutral Higgs bosons. For the measurement of trilinear 
couplings, we consider the multiple Higgs production processes. 
We delineate the regions of MSSM parameter space where the 
trilinear couplings of the neutral Higgs bosons could be measured 
at a high energy electron-positron collider.}

\FullConference{38th International Conference on High Energy Physics\\
		3-10 August 2016\\
		Chicago, USA}

\begin{document}
\section{Introduction}
\vspace{-.4cm}
The ATLAS and CMS experiments at the CERN Large Hadron Collider~(LHC) 
have discovered a particle 
with mass about $125$~GeV \cite{atlascms} whose properties are 
consistent with the Higgs boson of the standard model~(SM). 
To confirm that this particle is the Higgs boson of the SM, 
one must measure its couplings to other particles as well as to itself.
In the SM, the potential of the Higgs doublet $\phi$ which breaks
the $SU(2) \times U(1)$ gauge symmetry is  written as  \vspace{-0.25cm}
\be  V^h_{SM}=\lambda \left(\phi^2
-\frac{v^2}{2}\right)^2=\frac{m_h^2}{2}
 h^2 + \lambda_{hhh}^{SM} \frac{h^3}{3!}+\lambda_{hhhh}^{SM} 
 \frac{h^4}{4!} \quad ; \quad \phi=\frac{1}{\sqrt{2}}
(0 ~~ v+h)^T, \label{higgspotential}  \vspace{-0.25cm} \ee  
where  $h$ is the physical Higgs boson, and 
$\lambda_{hhh}^{SM}$  =  $3 m_h^2/m_Z^2$ = 5.6454 
(in units of $ (\sqrt{2} G_F)^{1/2} m_Z^2 $ = 33.77
GeV) and $\lambda_{hhhh}^{SM}  =  3m_h^2/m_Z^4$ = 0.00068 GeV$^{-2}$  
(in units of $(\sqrt{2} G_F)m_Z^4$ =1140.52 GeV$^2$) are the triple 
and  quartic Higgs  couplings, respectively. 
One must, therefore, measure these couplings in order to confirm
the idea of the spontaneous breaking of the underlying gauge symmetry.
Any deviation from the SM prediction for these couplings will signal
the existence of new physics beyond SM. 
Supersymmetry (SUSY)\cite{Nillesnath} is 
the leading candidate for the new physics  which  stabilizes the 
Higgs mass against large radiative corrections. The supersymmetric 
version of the SM, known as minimal supersymmetric 
standard model~(MSSM)\cite{Nillesnath}
not only solves the naturalness problem of the SM, but also leads to
gauge coupling unification at a large 
scale~($M_{\rm GUT} = 2 \times 10 ^{16}$~GeV).
Furthermore, the lightest supersymmetric particle, 
usually the lightest neutralino,  could be a possible 
dark matter candidate in $R$- parity conserving models. 
In the MSSM, there are two Higgs doublets~($H_2, H_1,$) with opposite 
hypercharge, which break the
SM gauge symmetry, and there are five physical Higgs 
bosons~($h^0, H^0, A^0, H^{\pm}$). In this paper we address the question
of the measurement of some of the trilinear
couplings of the neutral Higgs bosons
$h^0,  H^0$ at a high energy electron-positron~($e^+e^-$) collider.
\vspace{-.47cm}
\section{MSSM Higgs sector and trilinear couplings of $h^0, H^0$}
\begin{figure}
\vspace{-.35cm}
\begin{center}
\includegraphics[width=0.33\linewidth]{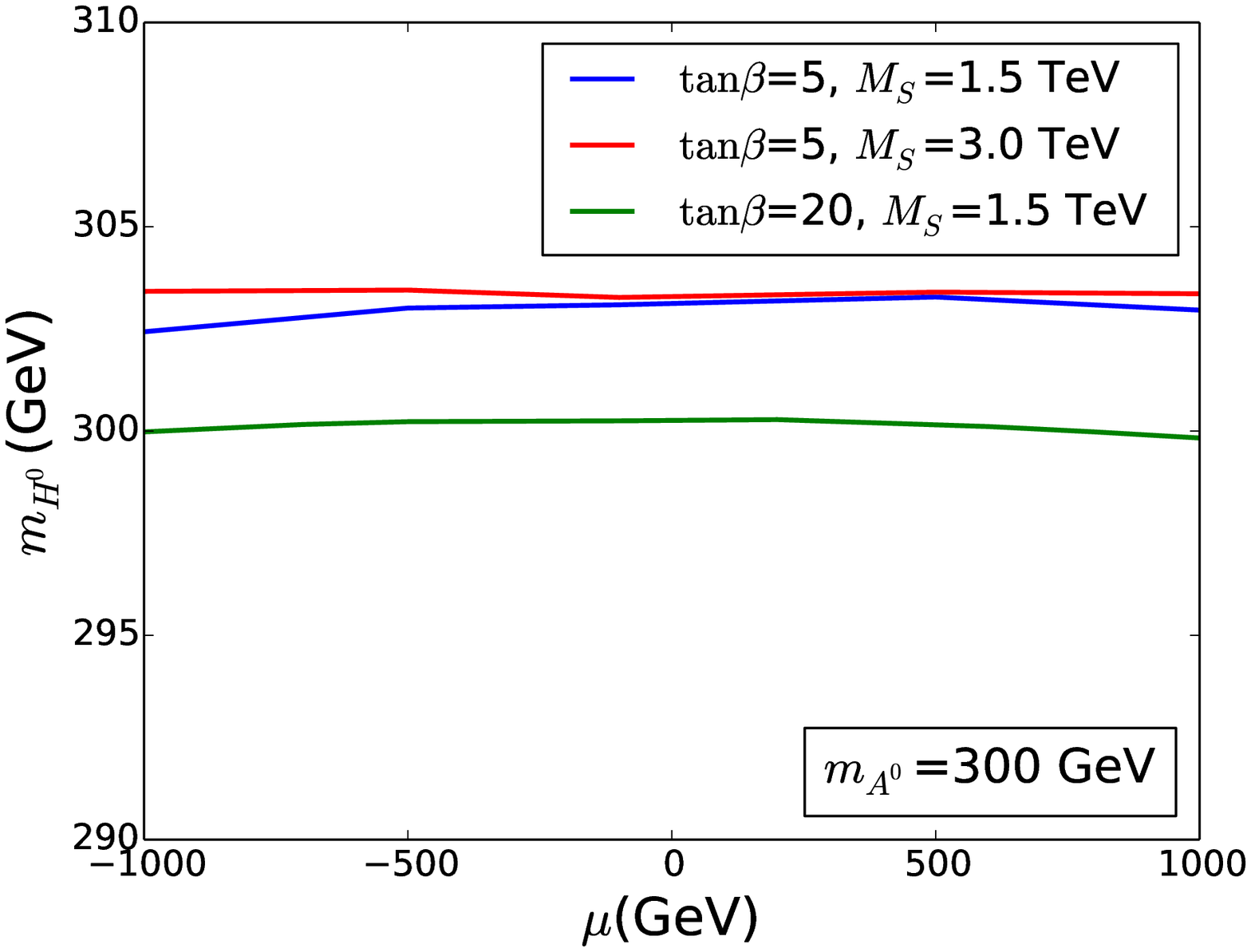}\vspace{-.1cm}\hspace{-.2cm}
\includegraphics[width=0.33\linewidth]{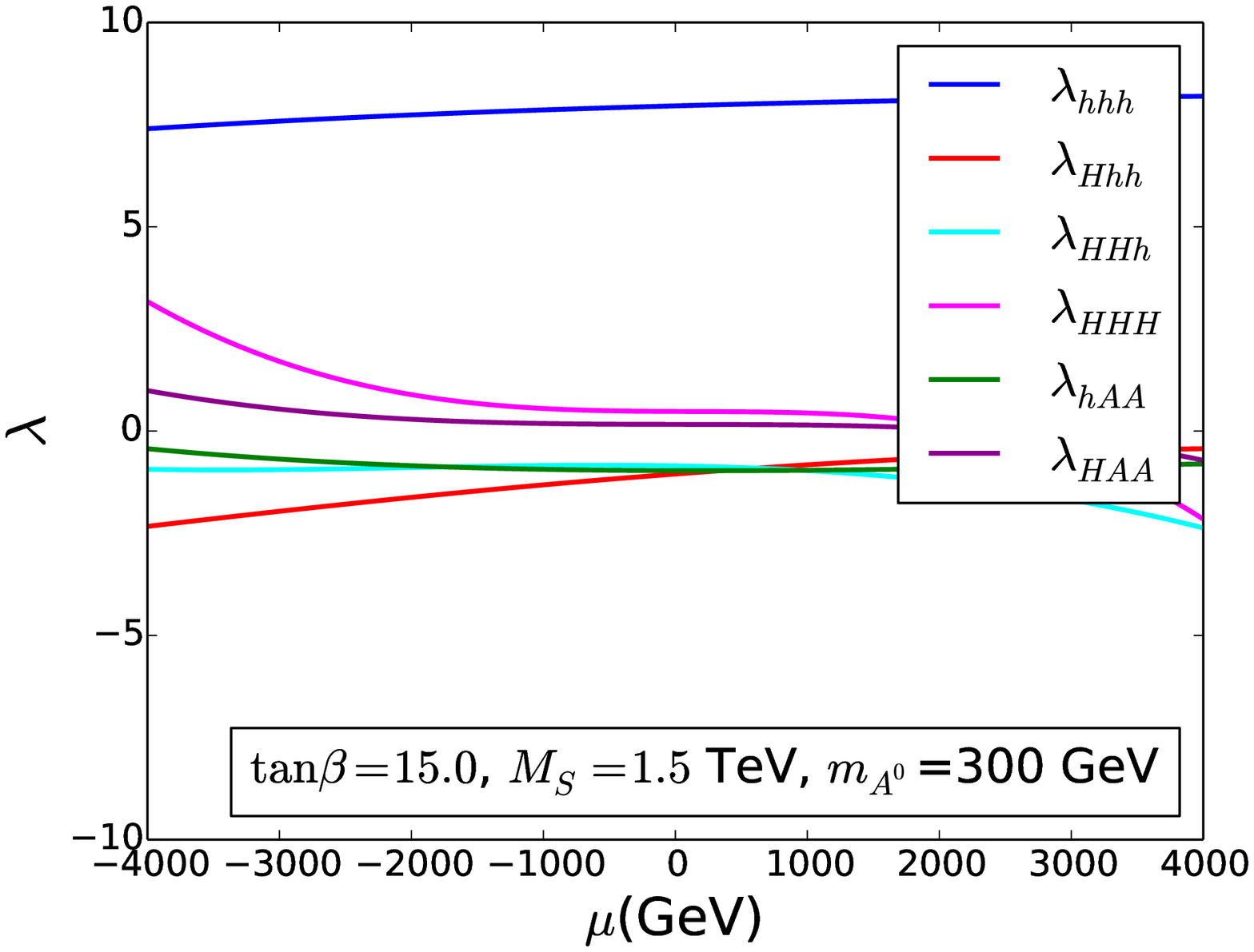}\vspace{-.1cm}\hspace{-.2cm}
\includegraphics[width=0.33\linewidth]{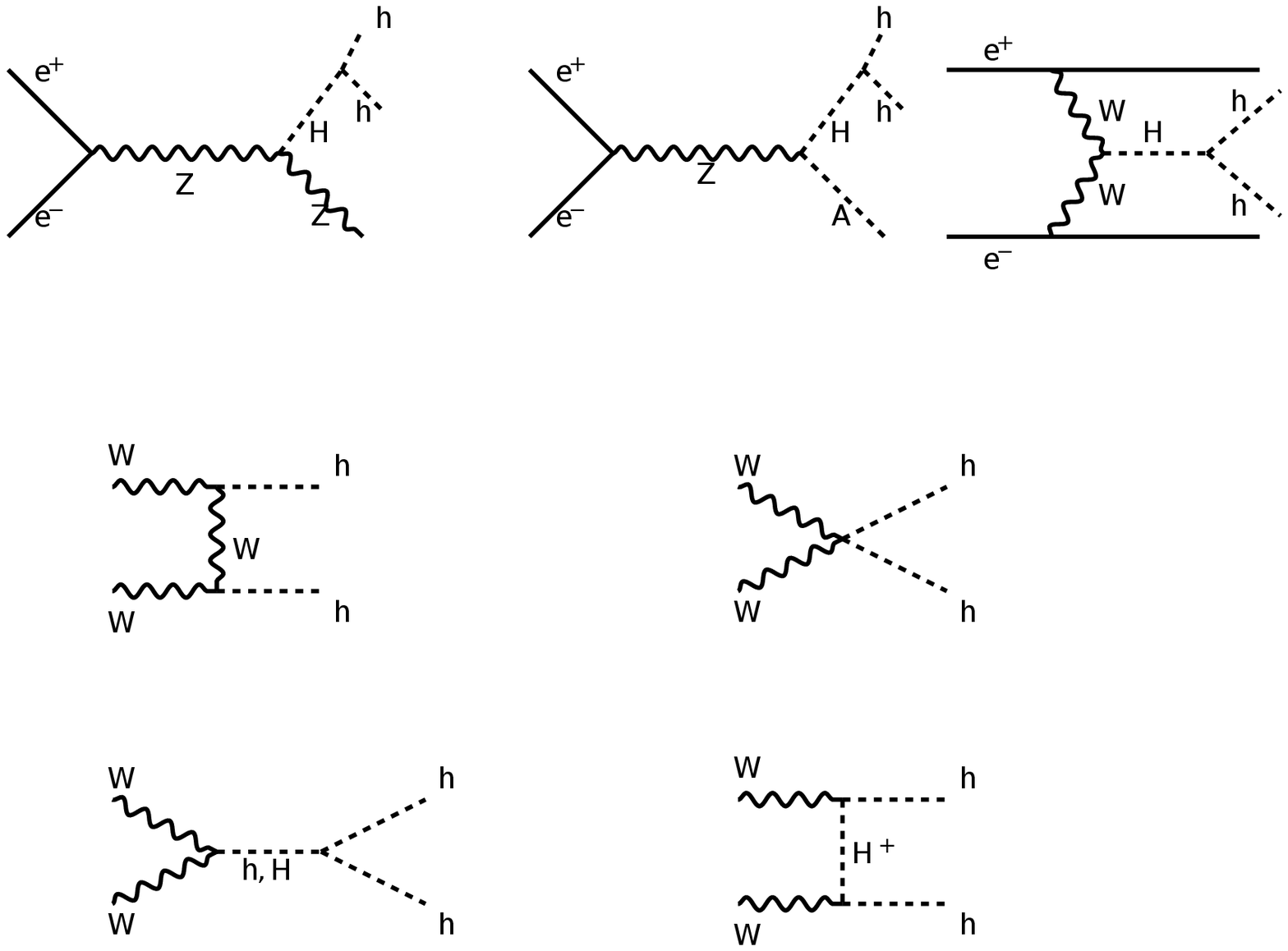}\vspace{-.1cm}
\caption{The mass of heavier Higgs boson $m_{H^0}$ as a function 
of $\mu$ parameter for fixed $\tan\beta$ and $M_S$(left panel); 
variation of radiatively corrected trilinear couplings 
in MSSM with $\mu$ parameter(central panel); Feynman diagrams for 
resonant production of  $hh$, through $e^+ e^-$ $\rightarrow$
$HZ$, $HA$, $\nu_e \bar\nu_e H $ ( where $H \rightarrow hh$ in the
final state) and through non-resonant $WW$ fusion (right panel).}
\label{higgsmass}
\end{center}
\vspace{-.88cm}
\end{figure}

\vspace{-.35cm}
In this section we shall 
consider the trilinear couplings of the CP-even 
Higgs bosons ($h^0,H^0$). At the tree level the masses of all the Higgs
bosons and their couplings can be written in terms of the mass of the CP-odd
Higgs boson~$m_{A^0}$ and $\tan\beta=<H_2^0>/<H_1^0>$. 
For the phenomenological survey, the Higgs sector of the MSSM 
can be divided into  two regions known as non-decoupling 
regime and decoupling regime. In the first case, $m_{A^0}$ $\le$ $130$~GeV, 
and heavy Higgs~($H^0$) is considered as SM-like (observed) state. 
On the other hand, decoupling regime is represented by $m_{A^0}$ $>$ $300$~GeV, 
and the  light Higgs boson~($h^0$) is identified with the 
observed $125$~GeV Higgs boson. We shall here focus
on the decoupling scenario.  The mass matrix for the CP-even Higgs bosons
of the MSSM can be written as  \vspace{-0.25cm}
\begin{eqnarray}
\label{Eq:M2} {\mathcal M}^2 &=&  \left[ \begin{array}{cc}
m_{A^0}^2 \sin^2 \beta + m_Z^2 \cos^2\beta &
-(m_Z^2 + m_{A^0}^2) \sin\beta \cos \beta\\
-(m_Z^2 + m_{A^0}^2) \sin\beta \cos \beta & m_{A^0}^2 \cos^2 \beta
+ m_Z^2 \sin^2\beta
\end{array} \right]  + \frac{3 g^2}{16 \pi^{2} m_W^2} \left[ \begin{array}{cc}
\Delta_{11} & \Delta_{12}\\
\Delta_{12} & \Delta_{22}
\end{array} \right],
\label{massmatrix1}\vspace{-0.25cm} \end{eqnarray}
where the radiative corrections 
$\Delta_{ij}$ are sensitive, besides other parameters, 
to the top-stop masses. Other parameters that enter the
calculations are the the Higgs(ino) parameter $\mu,$ and $\tan\beta.$
We have used CalcHEP~\cite{Belyaev} for the numerical calculations of the
Higgs spectrum.  Furthermore, we have adjusted the trilinear parameter $A_t$
so as to fix the mass of the lightest Higgs~($m_{h^0}$) 
in the range $ 122 - 128$~GeV. In Fig. \ref{higgsmass} (left panel) 
we  show the variation  
of the CP-even heavy Higgs mass~($m_{H^0}$) as a function of the parameter $\mu$
for fixed values of the supersymmetry breaking scale ($M_S$) and $\tan\beta.$ 
We see from this Fig. that the heavy Higgs mass is weakly dependent on 
$M_S$ for fixed value of $\tan\beta.$ Now coming to the Higgs self couplings,
assuming $CP$-conservation, we have six couplings amongst the neutral Higgs
bosons of the MSSM. These are written as $\lambda_{hhh}$, $\lambda_{Hhh}$,
$\lambda_{hAA}$, $\lambda_{HAA}$, $\lambda_{HHH}$ and $\lambda_{HHh}$.
As in the case of the masses of the Higgs bosons, these trilinear couplings
also obtain  significant radiative corrections~\cite{OPtricooup}. We can, then,
write these trilinear couplings generically 
as $\lambda = \lambda^0 + \Delta \lambda$ where $\lambda^0$ is the 
tree-level coupling and $\Delta \lambda$ are the radiative correction.
In this study we shall consider only two of these trilinear couplings, namely
(in units of $(\sqrt{2}G_F)^{1/2}m_Z^2$) 
$ \lambda_{hhh}^0  =  3 \cos2\alpha \sin(\beta
+ \alpha)$ and $\lambda_{Hhh}^0  =  2\sin2\alpha \sin(\beta + \alpha) - \cos
2\alpha \cos(\beta + \alpha)$, where $\alpha$ is the mixing angle in the CP-even 
Higgs sector. These couplings are sensitive to $m_{A^0}$ 
values of upto $500$~GeV. The dependence  of the trilinear couplings 
on the parameter $\mu$ is shown in Fig.\ref{higgsmass}.
%___________________________________
%_______________________________________
\vspace{-.54cm}
\section{Measurement of trilinear couplings}
\begin{figure}
\vspace{-.3cm}
\begin{center}
\includegraphics[width=0.32\linewidth]{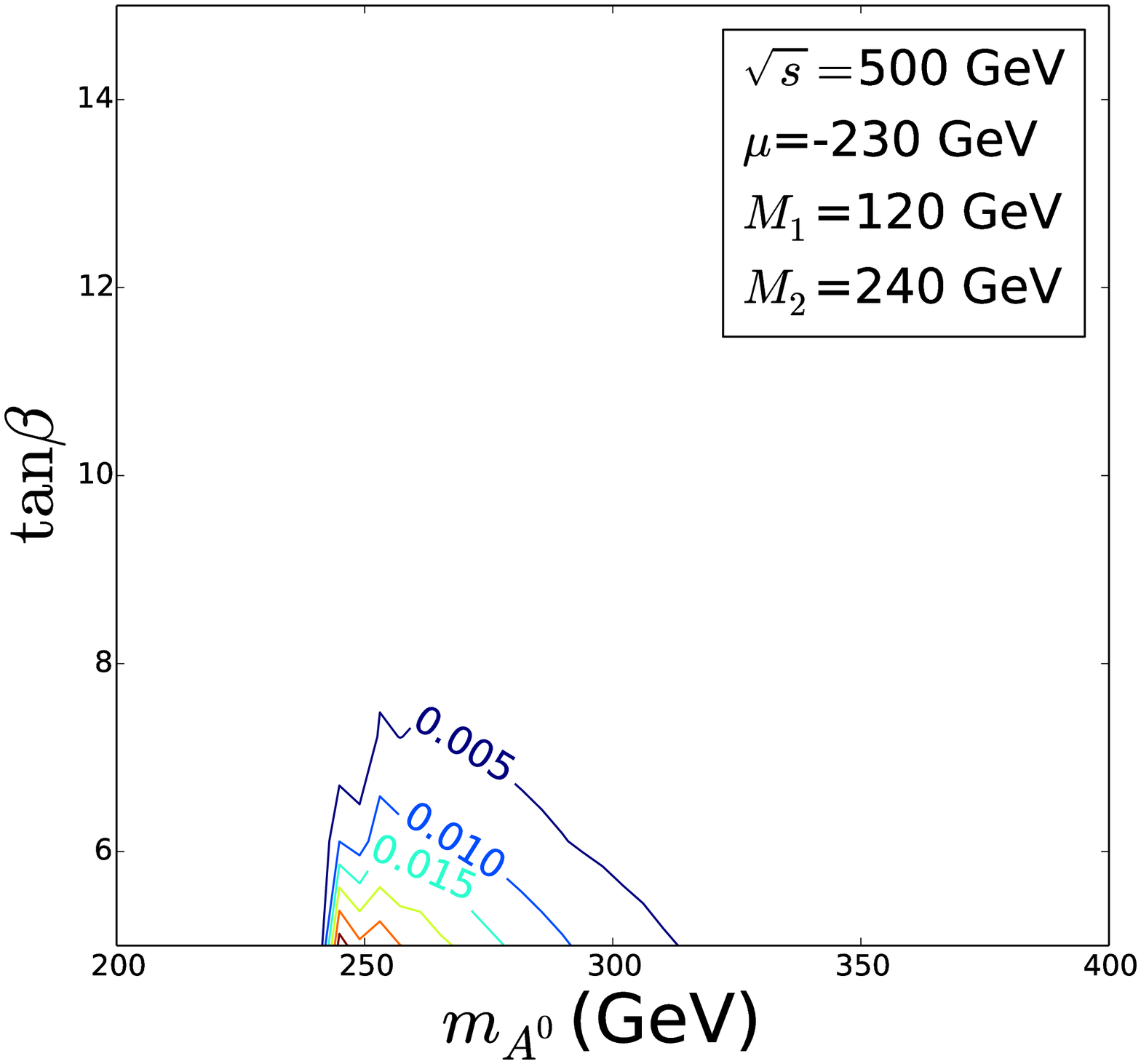}\vspace{-.1cm}
\includegraphics[width=0.32\linewidth]{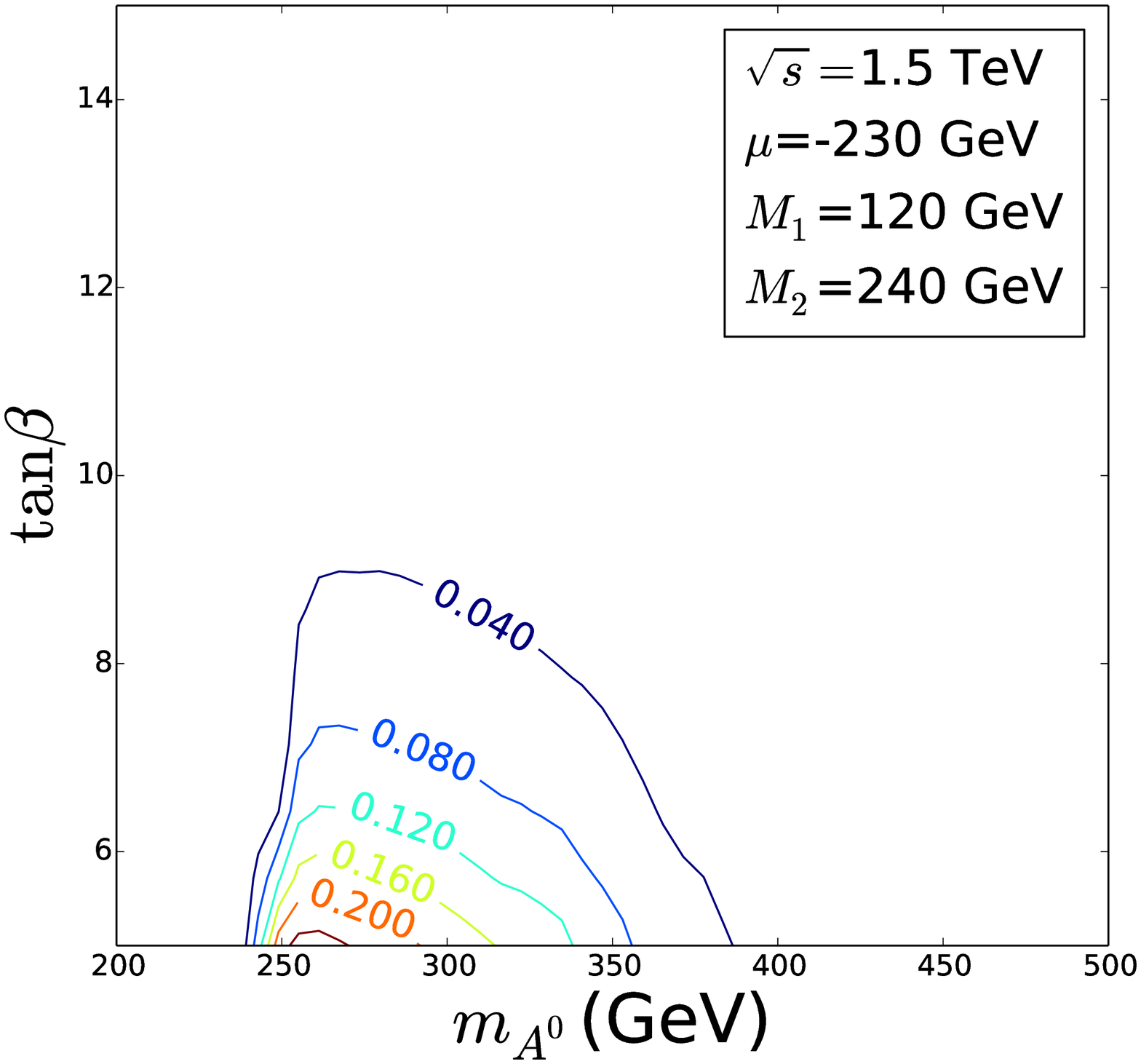}\vspace{-.1cm}
\includegraphics[width=0.32\linewidth]{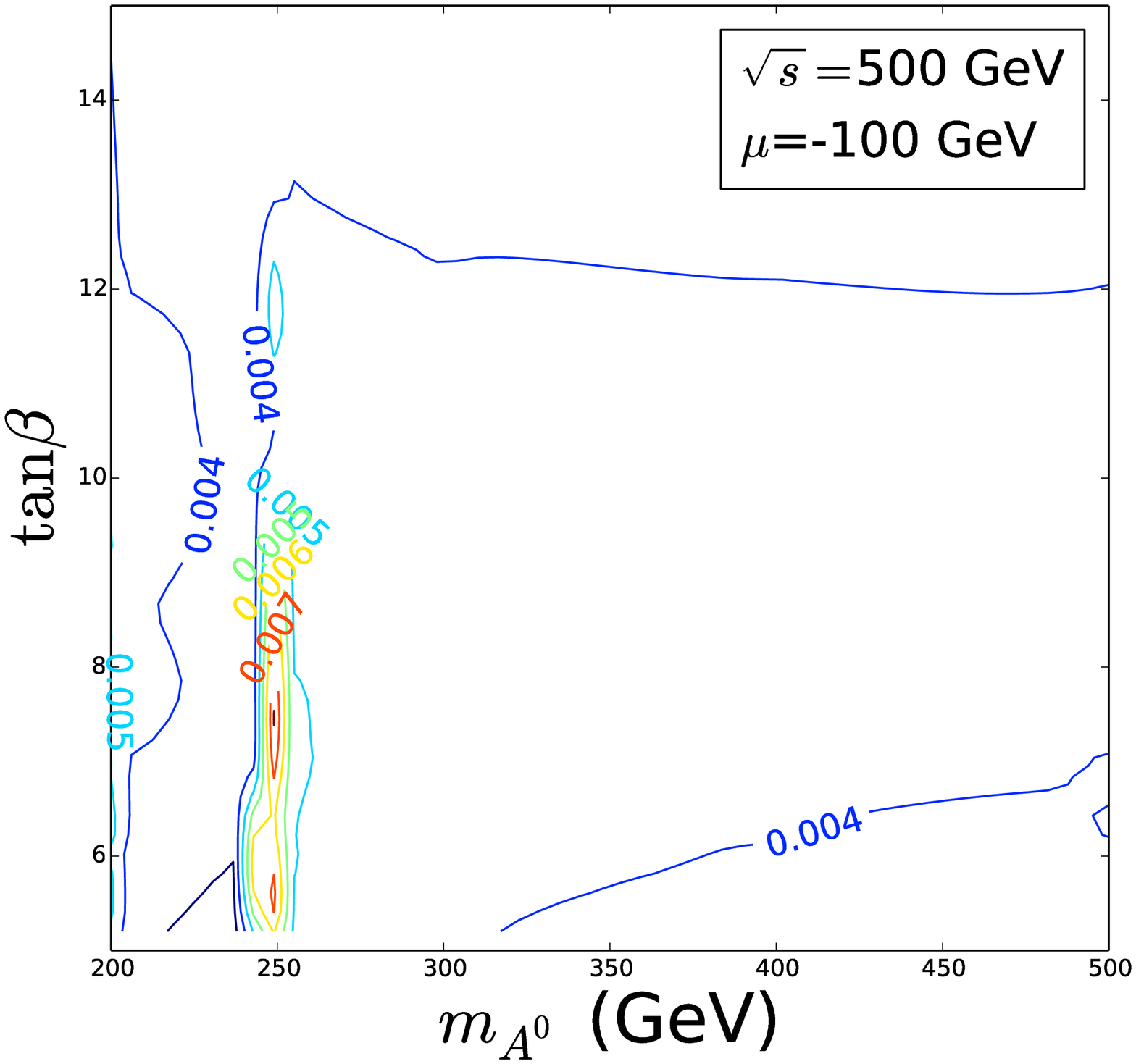}\vspace{-.1cm}
\caption{The contours of constant 
$\sigma(H) \times BR(H \rightarrow hh)$ (in fb)
(left and central panel); 
the contours of non-resonant $\sigma(ee \rightarrow hh \nu \bar\nu) $ (in fb)
via non-resonant WW fusion (right).}
\label{mssmplane}
\end{center}
\vspace{-1.40cm}
\end{figure}

\vspace{-.3cm}
We now discuss the multiple Higgs production processes which 
can be used to study the trilinear couplings involving $H^0$ and $h^0.$ 
We consider the production of heavy Higgs boson through 
Higgs-strahlung $e^+ e^- \rightarrow ZH $, associated production
with CP-odd Higgs boson $e^+ e^- \rightarrow AH $, and $WW$ fusion
mechanism $e^+ e^- \rightarrow \nu_e \bar\nu_e H $ 
(see Fig. \ref{higgsmass} (right panel) for 
Feynman diagrams and \cite{OPtricooup} for cross-sections). Multiple
light Higgs bosons ($h^0$) can be  produced through heavy CP-even
Higgs boson decay. Non- resonant $hh$ pair could also be 
produced with $Z$ or $A$ \cite{OPtricooup}. 
Associated production of $A$ with $h$, where $A$ decays to $hZ$, 
contributes as the background to the multiple Higgs production processes.
%__________________________
\begin{figure}
\vspace{-.3cm}
\begin{center}
\includegraphics[width=0.33\linewidth]{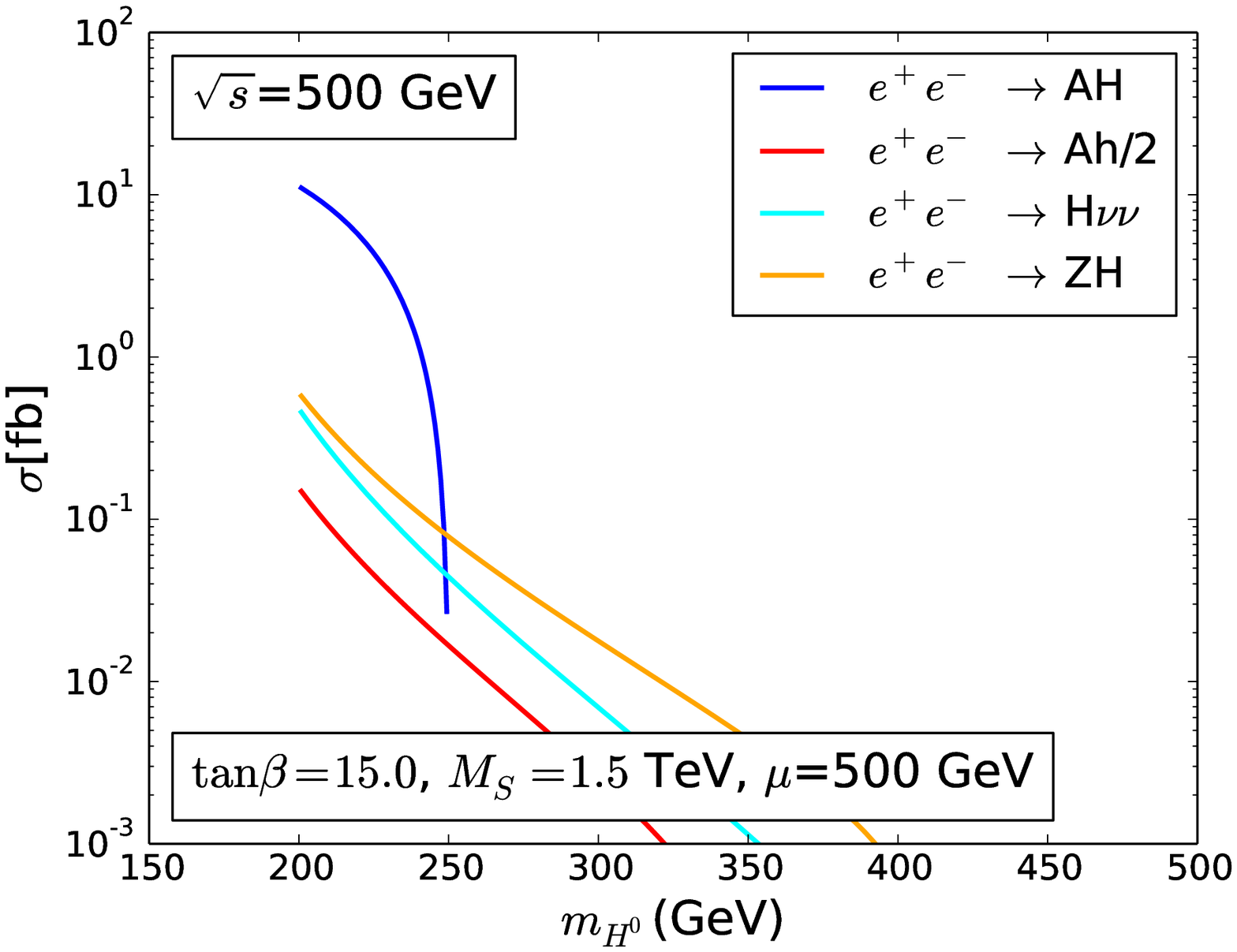}\hspace{-.2cm}
\includegraphics[width=0.33\linewidth]{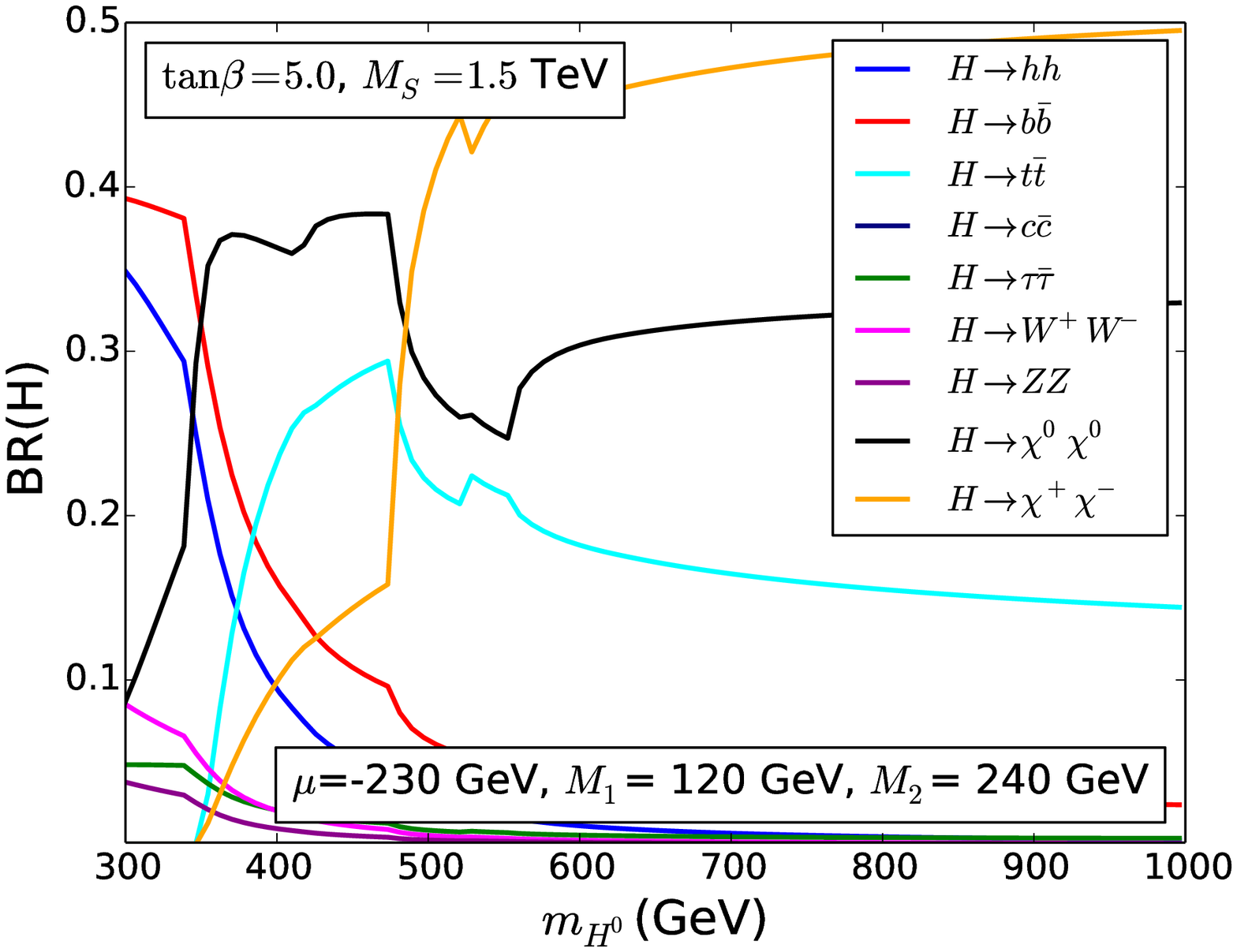}\hspace{-.2cm}
\includegraphics[width=0.33\linewidth]{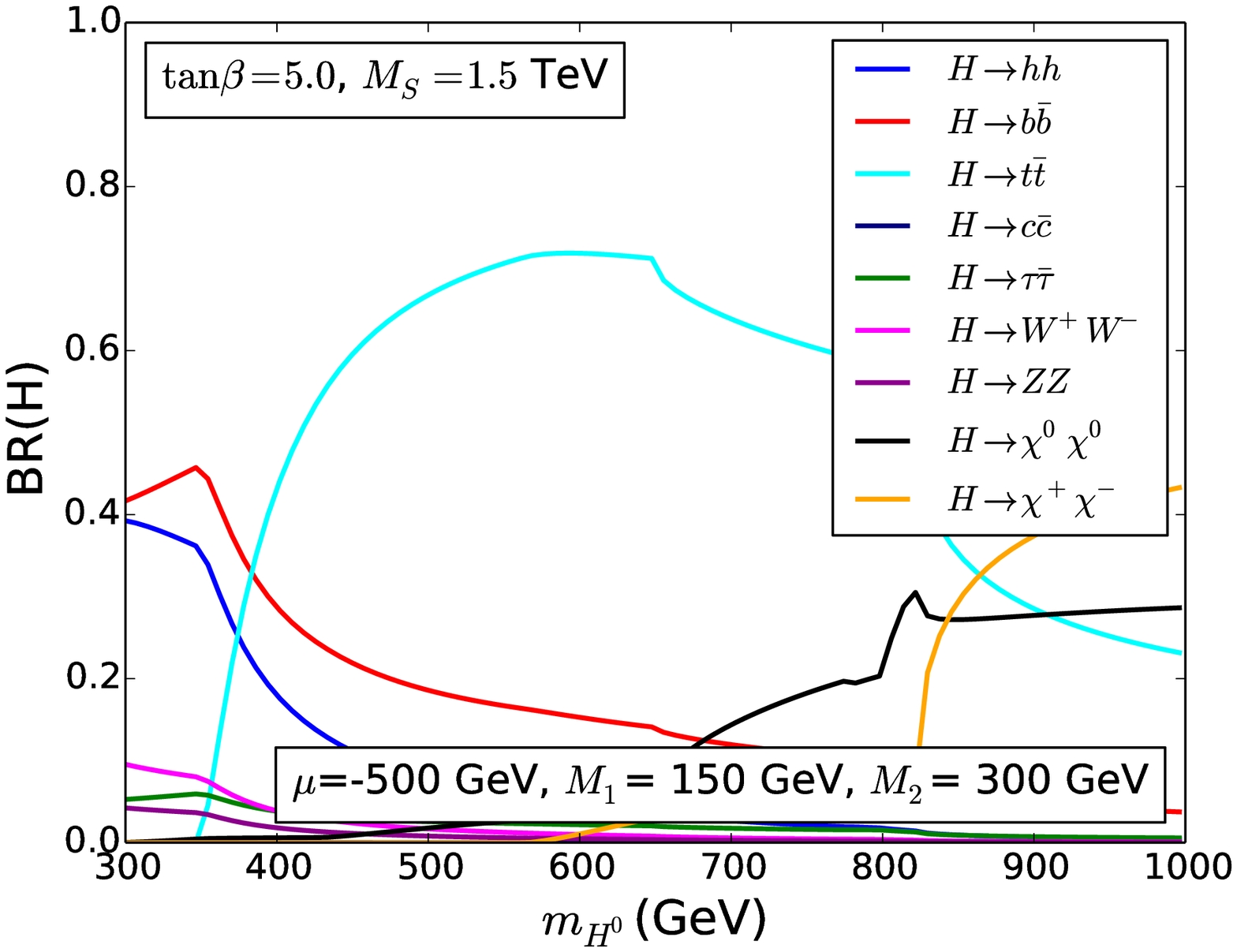}
\caption{Cross-section for $e^+ e^-$ $\rightarrow$ $AH,ZH,Ah/2,H\nu\bar\nu$ as 
the function of $m_{H^0}$ for $\sqrt{s}$ = 500 GeV (left); the branching 
fractions for different decay channels of $H$ for both the 
benchmark points(central and right).}
\label{crossection}
\end{center}
\vspace{-.74cm}
\end{figure}
%__________________________________
In  Fig. \ref{crossection}, left panel, we show the cross-section for
$e^+$ $e^-$ $\rightarrow$ $ZH$, $AH$, $H\nu\bar\nu$ as a function of $m_{H^0}$ 
for $\sqrt{s}$ = 500 GeV. We choose two benchmark values for the parameters 
($\mu,M_1,M_2$) = $(-230, 120, 240)$ and $(-500,150,300)$, respectively, 
so as to include SUSY particles in the final state of the heavy Higgs boson 
decay. In   Fig. \ref{crossection}, central and right panel,
we show the branching fractions of heavy 
Higgs decay into different channels. Second 
benchmark values have 
larger BR($H \rightarrow hh$) because neutralino and chargino 
spectrum is heavy as compared to the first benchmark point.
Consequently,   BR$(H \rightarrow \chi^0\chi^0,\chi^+\chi^-)$ 
is suppressed. In Fig. \ref{mssmplane}, for the 
measurement of $\lambda_{Hhh}$ coupling, we show the contours of 
$\sigma(H)$ $\times$ BR($H \rightarrow hh$) for $\sqrt{s}$ = $500$~GeV
 and $\sqrt{s}$ = $1.5$~TeV, respectively for the first benchmark point. 
The decay $H \rightarrow hh$ is kinematically forbidden for 
$m_{A^0}$ $\approx$ $m_{H^0}$ $\le$ $250$~GeV,  
and this branching ratio decreases as we move diagonally upward 
in the $(m_{A^0},\tan\beta)$ plane. The lower left corner 
of $(m_{A^0},\tan\beta)$ plane is the suitable 
region to measure $\lambda_{Hhh}$ coupling. The non-resonant WW fusion 
to $hh\nu_e\bar\nu_e$ final state involves both $\lambda_{hhh}$ and 
$\lambda_{Hhh}$ couplings. Having an estimate of $\lambda_{Hhh}$, we 
use this process to measure $\lambda_{hhh}$ coupling. We can see from 
the Fig. \ref{mssmplane}, right panel, 
that the $\sigma(hh\nu_e\bar\nu_e)$ is 
almost independent of $m_{A^0}$ and $\tan\beta$.
%_______________________________________
%________________________________________________
\section{Conclusions} \vspace{-0.42cm}
We have studied the trilinear couplings of the CP-even
Higgs bosons~($H^0, h^0$) in the minimal supersymmetric standard model,
and their measurement at a high energy $e^+ e^-$ collider.
In doing so we have identified the resonance observed at the CERN LHC
at $125$~GeV with the lightest CP even Higgs boson of the MSSM. By 
estimating the production cross-section 
of the  various processes involving multiple 
Higgs bosons, we have  delineated the regions 
in the ($m_{A^0}$,$\tan\beta$) space where trilinear 
couplings $\lambda_{Hhh}$ and $\lambda_{hhh}$ can be measured at 
an $e^+$ $e^-$ collider. For the coupling $\lambda_{Hhh}$, we have 
$\sigma (H) \times BR(H \rightarrow hh)$ 
$\approx$ $0.005$~fb  and $0.04$~fb for $\sqrt{s}$ = $500$~GeV 
and $\sqrt{s}$ = $1.5$~TeV, 
respectively, and it decreases along the diagonal upward direction 
in the $(m_{A^0},\tan\beta)$ plane. We estimate the 
coupling $\lambda_{hhh}$  through non-resonant WW fusion process. Precise 
knowledge of neutralino and chargino masses is
crucial in order to determine  the trilinear couplings of 
$H^0, h^0$ of the MSSM.

\end{document}